%Paper: astro-ph/9303020
%From: pyne@cfa160.harvard.edu (Ted Pyne)
%Date: Wed, 31 Mar 93 13:45:30 EST

\hoffset=0.7cm \voffset=0.2cm
\vbadness=10000
%
% The following statements redefine the basic fonts
% to be magnified by a factor 1.2
%

\font\sss=cmssq8 scaled 1000
\font\bf=cmbx10 scaled 1200
\font\bb=cmbx10 scaled 1920

\font\rs=cmr8 scaled 1200
\font\it=cmti10 scaled 1200

\font\sc=cmcsc10 scaled 1200
\font\tenrm=cmr10 scaled 1200
\font\sevenrm=cmr7 scaled 1200
\font\fiverm=cmr5 scaled 1200
\font\teni=cmmi10 scaled 1200
\font\seveni=cmmi7 scaled 1200
\font\fivei=cmmi5 scaled 1200
\font\tensy=cmsy10 scaled 1200
\font\sevensy=cmsy7 scaled 1200
\font\fivesy=cmsy5 scaled 1200

\font\tenbf=cmbx10 scaled 1200
\font\sevenbf=cmbx7 scaled 1200
\font\fivebf=cmbx5 scaled 1200
\font\tensl=cmsl10 scaled 1200
\font\tentt=cmtt10 scaled 1200
\font\tenit=cmti10 scaled 1200
\catcode`\@=11
\textfont0=\tenrm \scriptfont0=\sevenrm \scriptscriptfont0=\fiverm
\def\rm{\fam\z@\tenrm}
\textfont1=\teni \scriptfont1=\seveni \scriptscriptfont1=\fivei
\def\mit{\fam\@ne} \def\oldstyle{\fam\@ne\teni}
\textfont2=\tensy \scriptfont2=\sevensy \scriptscriptfont2=\fivesy
\def\cal{\fam\tw@}
\textfont3=\tenex \scriptfont3=\tenex \scriptscriptfont3=\tenex
\newfam\itfam \def\it{\fam\itfam\tenit} % \it is family 4
\textfont\itfam=\tenit
\newfam\slfam  % \sl is family 5
\textfont\slfam=\tensl
\newfam\bffam \def\bf{\fam\bffam\tenbf} % \bf is family 6
\textfont\bffam=\tenbf \scriptfont\bffam=\sevenbf
\scriptscriptfont\bffam=\fivebf
\newfam\ttfam  % \tt is family 7
\textfont\ttfam=\tentt
\catcode`\@=12
%
% This ends font redefinitions
%
\rm

% Definitions added in my version

\abovedisplayskip=30pt plus 4pt minus 10pt
\abovedisplayshortskip=20pt plus 4pt
\belowdisplayskip=30pt plus 4pt minus 10pt
\belowdisplayshortskip=28pt plus 4pt minus 4pt
\def\folio{\ifnum\pageno=1\nopagenumbers\else\number\pageno\fi}

\def\--{\! - \!}

\hfuzz=10pt \overfullrule=0pt
\vsize 8.75in
\hsize 6in
\baselineskip=14pt

\parindent 20pt \parskip 6pt

\def\blankline{\par\vskip \baselineskip}
\def\twocol#1{\halign{##\quad\hfil &##\hfil\cr #1}}

 \mathcode`*="002A

\def\_{\vrule height 0.8pt depth 0pt width 1em}

\newbox\grsign \setbox\grsign=\hbox{$>$} \newdimen\grdimen \grdimen=\ht\grsign
\newbox\simlessbox \newbox\simgreatbox
\setbox\simgreatbox=\hbox{\raise.5ex\hbox{$>$}\llap
     {\lower.5ex\hbox{$\sim$}}}\ht1=\grdimen\dp1=0pt
\setbox\simlessbox=\hbox{\raise.5ex\hbox{$<$}\llap
     {\lower.5ex\hbox{$\sim$}}}\ht2=\grdimen\dp2=0pt

\def\dot#1{\vbox{\baselineskip=-1pt\lineskip=1pt
     \halign{\hfil ##\hfil\cr.\cr $#1$\cr}}}
\def\ddot#1{\vbox{\baselineskip=-1pt\lineskip=1pt
     \halign{\hfil##\hfil\cr..\cr $#1$\cr}}}

\def\cfalh{\par\vfil\eject \vskip -12pt \moveleft 0.5in\vbox{
     \twocol{{\bb Center for Astrophysics}\hbox to 1.5in{} &\cr
     {\sss 60 Garden Street} & {\sss Harvard College Observatory} \cr
     {\sss Cambridge, Massachusetts 02138} & {\sss Smithsonian
          Astrophysical Observatory}\cr}}\par\blankline}

\def\listitem{\par \hangindent=50pt\hangafter=1
     $\ $\hbox to 20pt{\hfil $\bullet$ \hfil}}
\def\date#1{\par\hbox to \hsize{\hfil #1\qquad}\par}

\def\ref#1{$^{#1}$}
\def\title#1\endtitle{\par\vfil\eject
     \par\vbox to 1.5in {}{\bf #1}\par\vskip 1.5in\nobreak}
\def\author#1\endauthor{\par{\sc #1}\par\blankline}
\def\institution#1\endinstitution{{\rs #1}}

\def\section#1\endsection{\par\vfil\eject{\bf #1}\par\vskip 12pt\nobreak\rm}
\def\subsection#1\endsubsection{\vskip 14pt plus 50pt {\rm #1}\par
     \nobreak\blankline\nobreak\rm}

%\doublespace
\centerline{\bf Null Geodesics in Perturbed Spacetimes}
\vskip0.25truein
\centerline{Ted Pyne and Mark Birkinshaw}
\centerline{Harvard-Smithsonian Center for Astrophysics}
\centerline{60 Garden St., Cambridge, MA 02138}
\vskip0.5truein

\centerline{\bf ABSTRACT}

We present a generalization and refinement of the Sachs-Wolfe
technique which unifies many of the approaches taken to date and
clarifies both the physical and the mathematical character of the
method. We illustrate the formalism with a calculation of the behavior
of light passing a moving lens on a Minkowski background.

\noindent
{\it Subject headings:} gravitation---cosmology: cosmic microwave
background---cosmology: gravitational lensing

\vskip0.25truein
\noindent
{\bf I. Introduction}

A perfectly homogeneous and isotropic cosmic microwave background
radiation (CMBR) is possible only in a
Friedmann-Lema\^{\i}tre-Robertson-Walker (FLRW) spacetime
(Ehlers, Geren and Sachs, 1968).
For this reason, the extreme isotropy of the CMBR
has often been taken to indicate that our Universe is, in some sense,
well approximated by one of the FLRW models. Nevertheless,
there is structure in the Universe that must induce
distortions into the CMBR.

In the usual scenarios,
the structure that we observe today evolves via gravitational
instability from small inhomogeneities of the matter fields at the epoch of
matter-radiation equality, which occurs at temperature
$T_{EQ}\sim 5.5
\Omega_0 h^2$ eV (where $h=H_0/100\,{\rm kms}^{-1}\,{\rm Mpc}^{-1}$),
when the formation of galaxies and clusters by
gravitational collapse begins.
(Peebles and Yu, 1970; Bond and Efstathiou,
1984; Bond and Efstathiou, 1987; Suto, Gouda, and Sugiyama, 1990).
An important prediction of gravitational instability models
was first noticed by Sachs and Wolfe (1967)
who showed that even small amplitude
perturbations of FLRW spacetimes at recombination
(temperature $T_{REC}\sim .31$ eV, notably
close to the epoch of matter-radiation equality) cause appreciable
temperature fluctuations in the CMBR.

There are two main problems associated with a
Sachs-Wolfe calculation; relating the metric
perturbations to interesting physical perturbations of the matter
fields, and determining the
behavior of null geodesics in the perturbed spacetime.
Solving the first problem amounts to constructing the physical model
to be investigated. The behavior of electromagnetic radiation in the
model can then be understood in the geometric optics limit by
examining the behavior of null geodesics (Jordan, Ehlers, and Sachs,
1961).
The original work of Sachs and Wolfe (1967) concerned itself with
a particular perturbed Einstein de-Sitter spacetime and with a
single observable property of its null rays,
the redshift. This is an important
quantity because it is directly, and simply, related to the
temperature profile of a thermal radiation field such as the CMBR.
Much subsequent research
has been devoted to constructing perturbations about Einstein
de-Sitter spacetime which embody specific physical characteristics
(e.g. spheroidal perturbations) and using
the equivalent of the Sachs-Wolfe formula appropriate for the specific
perturbations in order to
understand the temperature pattern of the background radiation in
these models (Grishchuk and Zel'dovich, 1978;
Linder, 1988a; Arg\"ueso and
Martinez-Gonzalez, 1989; Arg\"ueso, Martinez-Gonzalez, and Sanz,
1989).
A formula for the redshift of null geodesics valid for perturbations
about the curved FLRW models as well as the flat was produced by Anile
and Motta in 1976.

Since gravitational lensing may be thought of as an aspect of the
theory of perturbed null geodesics, some researchers have used
Sachs-Wolfe like calculations to investigate the bending of light
rays. Linder has considered the spatial components of the
perturbed geodesic equation on an Einstein de-Sitter background,
in addition to the timelike component, and
has analyzed questions usually found in the realm of
gravitational lens theory with a Sachs-Wolfe type calculation
(Linder, 1988b, 1990). Martinez-Gonzalez, Sanz, and Silk (1990)
have exploited the fact that, under certain assumptions, metric
perturbations of the flat FLRW spacetime may be expressed by a small,
scalar potential which obeys an expanding-space analog of the usual
Poisson equation. They use this result to write formulae for both
redshift and deflection angle in physically transparent forms.

A complete generalization of the Sachs-Wolfe formalism should
solve for the timelike and spacelike components of null geodesics in a
metric perturbed spacetime of arbitrary background. Such a
generalization is presented here for the first time.
When the background is taken to be Einstein-de Sitter our
technique agrees with that of Sachs and Wolfe for the expression of
the photon redshift and with that of Linder (1988b) for the
perturbed photon path. When the
background is taken to be a curved FLRW spacetime,
our results agree with those of Anile and Motta for
the photon redshift (the quantity with which
they were concerned), but our method, unlike theirs, can also
be used to describe the spatial components of the perturbed photon
wavevector or lensing. Their formalism is inadequate for these issues
because the spatial components of their
primary equation (equation (7) of Anile and Motta, 1976),
do not take into account the distinct nature of the background
and perturbed paths to the correct order.

Since our formalism applies to general backgrounds it is able to
handle interesting cases which were not
previously amenable to a Sachs-Wolfe type calculation.
For example, our method may be used to
investigate perturbations of a Schwarzschild
background by long-wavelength gravitational waves.
We also expect that current developments in mathematical cosmology will
benefit from our more general formulation of the Sachs-Wolfe
technique. For instance, to date work has focused on models where
the density perturbations are small, in the sense that the fractional
density fluctuations $\delta\rho /\rho \ll 1$.
However, recent work by Futamase (1989) and Jacobs, Linder, and
Wagoner (1992) allows a description of a spacetime with {\it large}
density contrasts by a {\it small} metric perturbation. While the
background spacetime is not necessarily FLRW, ``clumpy''
cosmologies described in this way could be excellent models of our
Universe. In addition, the
notion of an inhomogeneous spacetime behaving, ``on average'', like an
FLRW spacetime is not rigorously understood and there is no reason to
believe that future models will behave
like linear perturbations of Einstein-de Sitter space.
We also hope that the technique presented here will be able to clarify
a difficult point in lensing theory, the nature of the distance
factors in the lens equation. Because of the difficulty in working
with the rigorous optical scalar equations (Sachs, 1961),
the relationship between
the formal mathematics of lensing and the, very successful, use of the
lens equation is not yet clear (Futamase and Sasaki, 1989;
Watanabe and Tomita, 1990).
The relationship of our technique to the
optical scalar equations is the subject of a forthcoming paper
(Pyne and Birkinshaw, work in progress).

This paper is organized as follows. In section II
we formulate an equation describing the
local separation of geodesics of the perturbed metric from geodesics of
the background. We then show that this equation is formally integrable
as a power series in the curvature of the background
spacetime.
In section III we use our solution to find the behavior of
light passing near a moving lens on a Minkowski background, obtaining
results in agreement with those obtained to date using other
approaches. In section IV we summarize our results.

\vskip0.25truein
\noindent
{\bf II. Null Geodesics in Perturbed Spacetimes}

{\bf (a) Conventions}.
In the following, Greek indices run over 0,1,2,3 while Latin indices
run over 1,2,3, the zeroth component is time-like, and
the summation convention is assumed. The metric signature is $+2$ and
we take $G=c=1$. The Riemann tensor convention is defined by equation
(A4).

{\bf (b) A Perturbed Jacobi Equation}.
Our starting point is the usual formalism of metric perturbation theory.
We consider the true metric to be the sum of a background metric
$g^{(0)}_{\mu\nu}$, and a small perturbation, $h_{\mu\nu}$,

$$ g_{\mu\nu}=g^{(0)}_{\mu\nu}+h_{\mu\nu}	.\eqno{(1)}$$

\noindent
We define $g^{\mu\nu}$ and $g^{(0)\mu\nu}$ by

$$ \eqalign{	g^{\mu\alpha}g_{\alpha\nu}	&=\delta^{\mu}_{\nu}
	\quad ,\cr
	g^{(0)\mu\alpha}g^{(0)}_{\alpha\nu}	&=\delta^{\mu}_{\nu}
\qquad .\cr}	\eqno{(2)}$$

\noindent
It is then a simple matter to verify that to order $h$

$$ g^{\mu\nu}=g^{(0)\mu\nu}-h^{\mu\nu}	\eqno{(3)}$$

\noindent
where $h^{\mu\nu}$ is defined by

$$ h^{\mu\nu}=g^{(0)\mu\alpha}g^{(0)\nu\beta}h_{\alpha\beta}
\qquad .\eqno{(4)}$$

\noindent
As usual (in a metric perturbation theory),
we raise and lower all tensor indices with the background metric.

The Levi-Civita connection of $g_{\mu\nu}$

$$\Gamma^{\mu}{}_{\alpha\beta}={1\over 2}g^{\mu\sigma}\left(
g_{\sigma\beta ,\alpha} + g_{\alpha\sigma ,\beta} -g_{\alpha\beta ,\sigma}
\right)	\eqno{(5)}$$

\noindent
can be split into zeroth and first order components in $h$

$$\Gamma^{\mu}{}_{\alpha\beta}=\Gamma^{(0)\mu}{}_{\alpha\beta}
+\Gamma^{(1)\mu}{}_{\alpha\beta}	\eqno{(6)}$$

\noindent
where

$$\Gamma^{(0)\mu}{}_{\alpha\beta}={1\over 2}g^{(0)\mu\sigma}\left(
g^{(0)}_{\sigma\beta ,\alpha}+g^{(0)}_{\alpha\sigma ,\beta}
-g^{(0)}_{\alpha\beta ,\sigma}	\right) \eqno{(7)}$$

\noindent
and

$$\eqalign{	\Gamma^{(1)\mu}{}_{\alpha\beta}	&={1\over 2}
g^{(0)\mu\sigma}\left( h_{\sigma\beta ,\alpha}+h_{\alpha\sigma ,\beta}
-h_{\alpha\beta ,\sigma}\right)		\cr
						&\qquad -{1\over 2}
h^{\mu\sigma}\left( g^{(0)}_{\sigma\beta ,\alpha} +g^{(0)}_{\alpha
\sigma ,\beta} -g^{(0)}_{\alpha\beta ,\sigma} \right)
\cr
	&={1\over 2}g^{(0)\mu\sigma}\left( h_{\sigma\beta ;\alpha}
+h_{\alpha\sigma ;\beta}-h_{\alpha\beta ;\sigma}\right).	\cr}
	\eqno{(8)}		$$

\noindent
The semicolon above and in what follows denotes covariant
differentiation with respect to the Levi-Civita
connection of $g^{(0)}_{\mu\nu}$. We note that
$\Gamma^{(1)\mu}_{\alpha\beta}$ is tensorial, at least in the
usual restricted sense of metric perturbation theory (that is, under
infinitesimal co-ordinate transformations).

Let $x^{(0)\mu}(\lambda )$ be
a geodesic of the background spacetime with $\lambda$ affine.
We will sometimes refer to $x^{(0)\mu}( \lambda)$ as the ``unperturbed
path''. It satisfies the geodesic equation in the unperturbed spacetime,
($\cdot =d/d\lambda$)

$$ {\ddot x^{(0)\mu} } + \Gamma^{(0)\mu}{}_{\alpha
\beta} \left( x^{(0)} \right) {\dot x^{(0)\alpha} }
{\dot x^{(0)\beta} } =0 	.\eqno{(9)}$$

\noindent
The notation $\Gamma^{(0)\mu}{}_{\alpha\beta}\left( x^{(0)}
\right)$ is meant to convey explicitly that the connection terms
are evaluated on the unperturbed path.

Now consider the expression

$$ x^{\mu}(\lambda )=x^{(0)\mu}( \lambda ) +x^{(1)\mu}
(\lambda)	\eqno{(10)}$$

\noindent
where $x^{(0)\mu}(\lambda)$ is the unperturbed geodesic.
At this point, both $x^{\mu}(\lambda )$
and $x^{(1)\mu}(\lambda )$ are unspecified and so equation (10) can be
considered to define either $x^{\mu}$ or $x^{(1)\mu}$
once the other is given. We will derive
conditions on $x^{(1)\mu}(\lambda)$ which will be
necessary and sufficient if $x^{\mu}(\lambda)$ is to be geodesic
in the perturbed spacetime.
In the following derivation, we will truncate three expressions by
discarding terms of quadratic order in the products of
$x^{(1)\mu}$ and ${\dot x^{(1)\mu}}$. We will often refer to this as
``working to first order.'' The consistency
conditions for the truncations made will be discussed at the end of
the derivation.

Differentiating equation (10) twice and using equation (9)
gives

$$ {\ddot x^{\mu}}=-\Gamma^{(0)\mu}{}_{\alpha\beta}
\left( x^{(0)} \right) {\dot x^{(0)\alpha}}
{\dot x^{(0)\beta}}+{\ddot x^{(1)\mu}}.
\eqno{(11)}$$

\noindent
On the other hand, if $x^{\mu}(\lambda )$ is to be an
affinely parameterized geodesic of the perturbed spacetime we must
have

$$\eqalign{ {\ddot x^{\mu}}	&=-\Gamma^{\mu}
{}_{\alpha\beta} \left( x\right) {\dot x^{\alpha}}
{\dot x^{\beta}}	\cr
	&=-\Gamma^{(0)\mu}{}_{\alpha\beta} \left( x\right)
\left( {\dot x^{(0)\alpha}}{\dot x^{(0)\beta}}
+2{\dot x^{(0)\alpha}}{\dot x^{(1)\beta}}
\right)	\cr
	&\qquad -\Gamma^{(1)\mu}{}_{\alpha\beta}\left( x\right)
{\dot x^{(0)\alpha}}{\dot x^{(0)\beta}}	,	\cr}
 \eqno{(12)}$$

\noindent
where we have used equations (6) and (10),
keeping terms only to first order.
Provided no singularities intervene (which we can ensure by keeping to
small enough neighborhoods and regular points),
the connection terms near the perturbed path, $x$,
may be expanded about their values on the
unperturbed path, $x^{(0)}$, as

$$\eqalign{ \Gamma^{(0)\mu}{}_{\alpha\beta}\left( x\right)
	&=\Gamma^{(0)\mu}{}_{\alpha\beta}\left( x^{(0)} \right)
+\Gamma^{(0)\mu}{}_{\alpha\beta ,\tau} \left( x^{(0)} \right)
x^{(1)\tau}+ \dots	\cr
	\Gamma^{(1)\mu}{}_{\alpha\beta}\left( x\right) 	&=
\Gamma^{(1)\mu}{}_{\alpha\beta}\left( x^{(0)}\right)+\dots	\qquad .\cr}
\eqno{(13)}$$

\noindent
Substituting these expansions into equation (12) gives

$$\eqalign{ {\ddot x^{\mu}}	&=-\Gamma^{(0)\mu}
{}_{\alpha\beta}\left( x^{(0)}\right) {\dot x^{(0)\alpha}}
{\dot x^{(0)\beta}} -\Gamma^{(1)\mu}{}_{\alpha\beta}
\left( x^{(0)}\right){\dot x^{(0)\alpha}}
{\dot x^{(0)\beta}} 		\cr
	&\qquad -2\Gamma^{(0)\mu}{}_{\alpha\beta}\left( x^{(0)}\right)
{\dot x^{(0)\alpha}}{\dot x^{(1)\beta}} -
\Gamma^{(0)\mu}{}_{\alpha\beta ,\tau}\left( x^{(0)}\right)
{\dot x^{(0)\alpha}}{\dot x^{(0)\beta}}x^{(1)\tau}
.\cr} \eqno{(14)}$$

\noindent
Comparing equations (14) and (11) we conclude that $x^{\mu}(\lambda )$,
defined by equation (10), will be an affinely parametrized
geodesic of the perturbed spacetime provided $x^{(1)\mu}(\lambda )$
satisfies the system of four coupled, second-order differential
equations

$$ \left( {d^2\over d\lambda^2} +A
{d\over d\lambda} +B\right)
x^{(1)} =f \qquad ,	\eqno{(15)}$$

\noindent
where the $4\times 4$ matrices $A$ and $B$ and the
four-vector $f$ are defined by

$$\eqalign{ A^{\mu}{}_{\alpha}	&=2\Gamma^{(0)\mu}{}_{\tau\alpha}
k^{(0)\tau}	\cr
	B^{\mu}{}_{\alpha}		&=\Gamma^{(0)\mu}{}_{\tau
\sigma ,\alpha}k^{(0)\tau}k^{(0)\sigma} 	\cr
	f^{\mu}				&=-\Gamma^{(1)\mu}{}_{\tau\sigma}
k^{(0)\tau}k^{(0)\sigma}	\qquad .\cr}	\eqno{(16)}$$

\noindent
Anticipating the interpretation of null geodesics as photon paths
we have written ${\dot x^{(0)}}$ as $k^{(0)}$.
A $4\times 4$ identity matrix
is the implied coefficient of the
second-order derivative in equation (15).
We are using the matrix notation as a convenient shorthand for
unambiguous summations, e.g $\left( Bx^{(1)}\right)^{\mu}=B^{\mu}
{}_{\alpha}x^{(1)\alpha}$. We emphasize that equation (15) is to
hold along some segment of the unperturbed path and generates
solutions for the separation, $x^{(1)}$, of the perturbed path
relative to the background path.

We still need to discuss where the above equation is actually valid,
that is, the appropriate consistency criteria for the approximations
made. We will use Futamase's (1988) goodness-of-fit parameters
for this purpose. Let the typical magnitude of a component of
$h$ relative to that of a component
of $g^{(0)}$ be written $\epsilon^2$. Write the typical scales of $h$
and $g^{(0)}$ as $l$ and $L$, respectively. Put $\kappa=l/L$.
Also, for simplicity (though without loss of generality),
assume that the initial conditions appropriate
for the sought-after solution are given at $\lambda=0$. Let $k^{(0)}$
denote the magnitude of a typical component of the unperturbed
wavevector at $\lambda=0$. The consistency criteria for the
approximations made above may then be found with the help of
equations (9), above, and (20) below to be $\epsilon^2\ll 1$,
$\epsilon^2\ll \kappa$, and

$$\epsilon^2 L\ln \left({\lambda k^{(0)}\over L} +1 \right)
\ll {\rm min}(l,L).			\eqno{(17)}$$

\noindent
This last condition may be replaced with the
stricter $\epsilon^2\lambda k^{(0)}\ll {\rm min}(l,L)$.

In appendix A we show that
equation (15) is simply a perturbation of the Jacobi equation of the
background spacetime.
In its most elegant form, then, our equation reads

$$\nabla^2_{k^{(0)}}x^{(1)}-R(k^{(0)},x^{(1)})k^{(0)}=f \eqno{(18)}$$

\noindent
where $\nabla$ is the Christoffel connection of the background
spacetime and $R$ is its curvature tensor (with implicit
superscript $(0)$). Remembering the tensorial
nature of $f$, the tensorial character of our equation is now
immediate. The physical meaning of the equation is also clear. The
geodesic of the perturbed metric differs from that of the background
because of
\item{(1)} a force-like term associated with the perturbation
(encoded in $f$) and
\item{(2)} the usual geodesic deviation induced by the background
spacetime.

{\bf (c) Comment on Initial Data}.
We now address the issue of what are appropriate initial data
for the perturbed Jacobi equation. We know that the geodesic
which we seek to construct is uniquely specified
by a point of that geodesic, $p$, and the tangent vector to the
geodesic at that point, $k_p$. As equation (10) indicates, we can construct a
segment of the desired geodesic locally about $p$ from any
geodesic of the background which (in the naive sense) passes close to
$p$ with tangent vector close to $k_p$. An important point,
however, is that the initial data specifying the desired geodesic does
not, in general, generate a null geodesic of the background because
$k_p$ is generally not null in both the
perturbed and background metrics. As we wish to
construct null geodesics of the full metric using a null geodesic of
the background, we conclude that the initial
condition $k^{(1)}(\lambda_p)=0$ is not
appropriate for our equation. However, in the regions far from
a localized perturbation (where, loosely, $h\rightarrow 0$)
this subtlety is entirely avoided.

It is instructive to examine this point in another way. Suppose that
we have chosen a null geodesic of the background and that
a solution, $x^{(1)}(\lambda)$, of the perturbed Jacobi equation
has been found. The
condition that the constructed geodesic be null, to
order $h$, in the perturbed spacetime is

$$\eqalign{g_{\mu\nu}(x)k^{\mu}k^{\nu}	&=h_{\mu\nu}
(x^{(0)})k^{(0)\mu}
k^{(0)\nu}+2g^{(0)}_{\mu\nu}(x^{(0)})k^{(0)\mu}k^{(1)\nu}	\cr
		&\qquad +2g^{(0)}_{\mu\nu ,\rho}(x^{(0)})x^{(1)\rho}
k^{(0)\mu}k^{(0)\nu}	\cr
					&=0 .\cr} \eqno{(19)}$$

\noindent
Making use of the co-ordinate invariance of scalar
quantities, we evaluate the last term on the RHS of the first equality
by choosing co-ordinates
adapted to $g^{(0)}$ at any given point so that
we can replace the ordinary derivative with the covariant
derivative of $g^{(0)}$. Metric compatibility then tells us that this term
vanishes. We are left with the condition that
everywhere along the background geodesic we must have

$$g^{(0)}_{\mu\nu}k^{(0)\mu}k^{(1)\nu}=-{1\over 2}h_{\mu\nu}
k^{(0)\mu}k^{(0)\nu}.				\eqno{(20)}$$

We now show that this condition can always be enforced.
We choose to work with co-ordinates for which the background
connection coefficients vanish along the unperturbed path.
In these co-ordinates, equation (15) implies

$$\eqalign{ {d\over d\lambda}\left( g^{(0)}_{\mu\nu}k^{(0)\mu}
k^{(1)\nu} \right)	&=g^{(0)}_{\mu\nu}k^{(0)\mu}\left(-
\Gamma^{(0)\nu}{}_{\alpha\gamma,\beta}k^{(0)\alpha}k^{(0)\gamma}x^{(1)\beta}
-\Gamma^{(1)\nu}{}_{\alpha\beta}k^{(0)\alpha}k^{(0)\beta}\right)	\cr
			&=-g^{(0)}_{\mu\nu}\Gamma^{(1)\nu}{}_{\alpha\beta}
k^{(0)\mu}k^{(0)\alpha}k^{(0)\beta}	\cr}\eqno{(21)}$$

\noindent
where the first term on the RHS vanishes as a consequence of the
identity

$$R^{(0)}{}_{\mu\alpha\beta\gamma}k^{(0)\mu}k^{(0)\alpha}
k^{(0)\gamma}x^{(1)\beta}=0	$$

\noindent
written out in the choosen co-ordinates.
The RHS of equation (21) may be shown, with the aid of equation (8),
to be

%$$=-{1\over 2}h_{\alpha\beta ,\mu}k^{(0)\mu}
%k^{(0)\beta}k^{(0)\alpha}+h_{\mu\rho}\Gamma^{(0)\rho}{}_{\alpha\beta}
%k^{(0)\mu}k^{(0)\alpha}k^{(0)\beta}	\eqno{(21)}$$
%
%\noindent
%or, equivalently,

$$=-{1\over 2}k^{(0)\mu}k^{(0)\alpha}k^{(0)\beta}
\left( h_{\alpha\beta ,\mu}-\Gamma^{(0)\rho}{}_{\alpha\mu}h_{\rho\beta}
-\Gamma^{(0)\rho}{}_{\mu\beta}h_{\alpha\rho} \right).\eqno{(22)}$$

\noindent
We have thus shown that

$${d\over d\lambda} \left( g^{(0)}_{\mu\nu}k^{(0)\mu}k^{(1)\nu}
\right) ={D\over d\lambda} \left( -{1\over 2}h_{\mu\nu}k^{(0)\mu}
k^{(0)\nu}\right)			\eqno{(23)}$$

\noindent
where $D/d\lambda$ is the covariant derivative of $g^{(0)}$ along
$x^{(0)}$.

On scalars we have $D/d\lambda =d/d\lambda$ so we can conclude
that if equation (20) is satisfied at any point along the background
geodesic it will be satisfied at every point. Notice that it is very
simple to impose equation (20) at the initial point but that, in
general, $k^{(1)}=0$ is not an acceptable solution because
at a given point a null vector in the
perturbed metric is not necessarily null in the background metric.
The calculation above thus amounts to a direct check that
$\langle k^{(0)}+k^{(1)} ,k^{(0)}+k^{(1)}\rangle_{g^{(0)}+h}$
is a constant of geodesic motion to order $h$,
as it must be if this procedure is
to be meaningful (the angle brackets denote the inner product
with respect to the metric of the subscript).

{\bf (d) A Solution to the Perturbed Jacobi Equation}.
Having found the domain for which our equation, (15), is valid
and having shown that the equation generates null paths from null paths,
we construct
a formal solution. The techniques we will use are simply the matrix
analogues of familiar methods for dealing with ordinary differential
equations. First we perform a change of variables in order to
eliminate the first derivative term in our equation. To accomplish this,
let $P(\lambda ,a)$ be a $4\times 4$ matrix function
of two real arguments, and
$v(\lambda )$ a vector such that $x^{(1)}=Pv$. Then in terms of $v$,
equation (15) becomes

$$ {\ddot v}+P^{-1}\left( 2{\dot P}+A P \right) {\dot v}
+P^{-1}\left( {\ddot P} +A {\dot P} +B P \right)v
=P^{-1}f					\eqno{(24)}$$

\noindent
where we have assumed $P$ non-singular (justified below).
Now choose $P$ to satisfy

$${\dot P}=-{A \over 2} P,			\eqno{(25)}$$

\noindent
The solution is the path-ordered exponential,

$$ P(\lambda,a) ={\cal P}\exp \left( -{1\over 2}\int_a^{\lambda}
A (\tau )\, d\tau \right)		\eqno{(26)}$$

\noindent
(a short introduction to the path-ordering symbol, $\cal P$,
is provided in appendix B). We remark that $P$ is exactly Synge's
parallel propagator (Synge, 1960). We will call $P$ the connector,
following De Felice and Clarke (1990).
Anticipating our final result, we
see that the effect of the change of variables above is basically to
untangle the co-ordinate basis, just as if we were working with a parallel
propagated tetrad frame.

In writing equation (26)
we have set $P(a,a)=1_d$, the four-dimensional identity matrix.
This initial condition and equation (25) can
be used to show the important identity

$$ P(\lambda ,\lambda_1 )P(\lambda_1 , a)=P(\lambda ,a) \eqno{(27)}$$

\noindent
and so

$$ P(\lambda ,a)^{-1}=P(a,\lambda )	\eqno{(28)}$$

\noindent
proving that $P$ is invertible.

With $P$ chosen in this manner, equation (24) now becomes

$${\ddot v}+P^{-1}\left( -\left( {A \over 2}\right)^2
-{ {\dot A}\over 2} +B \right) Pv=P^{-1}f \eqno{(29)}$$

\noindent
Written out explicitly using equation (16) the quantity in
parentheses reveals itself to be
$-R^{(0)\mu}{}_{\nu\rho\sigma}k^{(0)\nu}k^{(0)\rho}$,
which we will write as the matrix $-R$, using notation consistent with
(18).

We write (29) as a first-order system

$${d\over d\lambda}\pmatrix{v \cr {\dot v} \cr}-\pmatrix{0 & 1_d \cr
P^{-1}RP & 0 \cr}\pmatrix{v \cr {\dot v} \cr}=\pmatrix{0 \cr P^{-1}f
\cr}	\eqno{(30)}$$

\noindent
which can be solved by constructing the
transition matrix of the system,
$U(\lambda ,a)$, which is the solution of its homogeneous part
(Humi and Miller, 1988). This is the familiar method of Greens
function solution. We will use the terms transition matrix and Greens
function interchangeably in refering to $U(\lambda ,a)$.
Another path-ordering is needed to actually construct the
transition matrix, with the result

$$U(\lambda ,a)={\cal P}\exp \left( \int^{\lambda}_a
\pmatrix{0 & 1_d \cr P(\tau ,a)^{-1}R(\tau )P(\tau ,a) & 0\cr}
\, d\tau \right)		\eqno{(31)}$$

\noindent
where the boundary condition $U(a,a)=1_d$ has been imposed. As a
result, equations (27) and (28) remain valid with $U$ written
in place of $P$. Defining the column vectors $y=(v, {\dot v})$ and
$s=(0,P^{-1}f)$ the formal solution to equation (29) with initial value
boundary data is

$$  y(\lambda )=U(\lambda ,a)y(a)
+\int_a^{\lambda}U(\lambda ,\tau )s(\tau )\,d\tau
					\eqno{(32)}$$

\noindent
Confirmation that this expression solves equation (29) is
provided by differentiation.

While (32) is a formal solution to the perturbed Jacobi
equation (29; and hence 15),
at this point its utility is far from evident. In particular,
the presence of two path-ordered exponentials makes the
calculations difficult. Had we not made the initial change of
variables from $x^{(1)}$ to $v$ we could have solved the system
without recourse to the connector. We would then need only a single
path-ordering, that responsible for constructing the appropriate
Greens function. Appendix C outlines this approach.
We feel, however,
that the strength of the particular transcription of the solution
to the perturbed Jacobi equation given above, equation (32), lies
in its explicit geometric character. This is most evident by choosing
co-ordinates for which the Christoffel coefficients of the background
vanish along the background geodesic. In this case
equation (32) applies with $P=1_d$, $s=(0,f)$,
$y=(x^{(1)},k^{(1)})$ and

$$U(\lambda ,a)={\cal P}\exp \left( \int^{\lambda}_a
\pmatrix{0 & 1_d \cr R & 0 \cr}\, d\tau \right)	\eqno{(33)}$$

\noindent
clearly expressing the relevance of the background curvature to the
solution.

\vskip0.25truein
\noindent
{\bf III. The Moving Lens}

As an illustration of our method, we calculate the asymptotic behavior
of a light ray passing by a point mass
moving on a Minkowski background. Elements of this scenario have been
analyzed by Birkinshaw and Gull (1983) without a fully general-relativistic
treatment. Our formalism provides an easy and unified way
of handling the problem in the context of general relativity. This
particular case is especially simple because Minkowski space is flat,
allowing us to choose co-ordinates such that $A=B=0$, and hence allowing
equation (15) to be solved by direct integration. In appendix D we show
how the propagator techniques of section II can be used to gain an
equivalent solution.

{\bf (a) The Metric}.
Let primed co-ordinates denote the frame in which the lens is at rest
at the origin of co-ordinates.
In this frame the metric is of the usual Schwarzschild form,
$$ ds^{\prime 2}=-\left( 1-{2m\over r^{\prime}}\right) dt^{\prime 2}
+ \left( 1-{2m\over r^{\prime}} \right)^{-1} dr^{\prime 2}
+r^{\prime 2}d\theta^{\prime 2}+r^{\prime 2}\sin\theta^{\prime}
d\phi^{\prime 2}\qquad ,	\eqno{\rm (34)}$$
where $m$ is the mass of the lens.
Expanding to first order in $m /r^{\prime}$ we break the metric up
into a sum of the Minkowski metric in spherical co-ordinates
and a perturbation of the form
$$ h_{\alpha^{\prime}\beta^{\prime}}={2m\over r^{\prime}}\left(
\delta^0_{\alpha^{\prime}}\delta^0_{\beta^{\prime}} +
\delta^1_{\alpha^{\prime}}
\delta^1_{\beta^{\prime}}\right)	.\eqno{\rm (35)}$$
Transforming to Cartesian co-ordinates puts the background
Minkowski metric into its usual form, diag($-1,1,1,1$), and puts the
perturbation into the form
$$\eqalign{	h_{0^{\prime}0^{\prime}}\left( x^{\prime} \right)
				&={2m\over r^{\prime}}	\cr
	h_{0^{\prime}i^{\prime}}\left( x^{\prime} \right)
					&=h_{i^{\prime}0^{\prime}}
\left( x^{\prime} \right)=0	\cr
	h_{i^{\prime}j^{\prime}}\left( x^{\prime} \right)
					&={2m\over r^{\prime 3}}
x_{i^{\prime}}x_{j^{\prime}}	\cr}	\eqno{\rm (36)}$$
where
$$ r^{\prime}=\left[ x^{\prime 2}+y^{\prime 2}+z^{\prime 2}
\right]^{1/2}	.\eqno{\rm (37)}$$

\noindent
Note that the $x_{i^{\prime}}$ appearing in
equation (36) are not vectors but the primed co-ordinate
functions. In particular no index raising or lowering via a metric is
taking place, $x_{i^{\prime}}=x^{i^{\prime}}$. This remark will apply
equally well to the unprimed co-ordinate functions.

We are interested in the background metric and metric perturbation
in the frame in which the lens appears to be moving with velocity
$\underline{v}$, with the underbar
denoting a three-vector quantity.
We will use unprimed variables for this frame.
If we agree to work only to linear order in $v$ the connection between
our two sets of co-ordinates is simply a linearized Lorentz
transformation
$$\eqalign{	t^{\prime}	&=t-\underline{v}\cdot \underline{x}\cr
	\underline{x}^{\prime}	&=\underline{x}-
\underline{v}t .\cr} \eqno{\rm (38)}$$
where the dot product is a convenient
shorthand for summation of spatial indices, e.g. $\underline{v}
\cdot \underline{x}=v^ix_i$.
Under this transformation the background Minkowski metric is
unchanged while the metric perturbation becomes
$$\eqalign{	h_{00}	&={2m\over r}	\cr
		h_{0i}	&=h_{i0}=-{2m\over r}v_i -{2m\over r^3}
\left( \underline{v}\cdot\underline{x} \right) x_i	\cr
		h_{ij}	&={2m\over r^3}\left( x_ix_j -x_jv_it
	-x_iv_jt \right)	\cr}	\eqno{\rm (39)}$$
with
$$ r=\sqrt{ \left( \underline{x}-\underline{v}t \right)\cdot
\left( \underline{x}-\underline{v}t \right) }\ .\eqno{\rm (40)}$$

The global Galilean transformation, (38), may be used
to define the three-vector $\underline{v}$
because the background is Minkowski and because
products ${\cal O}(vh)$ are ignored, so that $v^i=v_i$.
Although this is not globally valid in general
relativity, the approximations adopted are consistent to the order
claimed, as may be discovered by repeating the calculation using
the weak-field equations of general relativity, where
the perturbation can be written ${\rm diag}(-2\phi,-2\phi,
-2\phi,-2\phi )$ with $\phi$ the Newtonian potential of the perturbation,
taken to be a moving point mass (see e.g. Weinberg, 1972).

{\bf (b) Connection Terms}.
Simple calculation using equations (8) and (39) produces
(to ${\cal O}(v,h)$),
$$\eqalign{
		\Gamma^{(1)0}_{00}	&=-{m\over r^3}\underline{v}\cdot
                                   \underline{x}	\cr
		\Gamma^{(1)0}_{0i}	&={m\over r^3}\left( x_i-v_it \right)
                                                        \cr
		\Gamma^{(1)0}_{jk}	&={2m\over r^3} \left( \underline{v}
                                  \cdot\underline{x} \right)
                                  \delta_{jk} -{m\over r^3}
                                  \left( v_jx_k+v_kx_j \right)
                                  -{3m\over r^5}
                                  \left( \underline{v}\cdot
                                  \underline{x} \right) x_jx_k
                                                    	\cr
                \Gamma^{(1)i}_{00}	&={m\over r^3}\left( x^i-v^it \right)
                                                        \cr
		\Gamma^{(1)i}_{0j}	&={3m\over r^5}\left( \underline{v}
                                  \cdot\underline{x}
                                  \right)x_ix_j +{m\over r^3}
                                  \left( x_jv_i-3x_iv_j \right)
                                                   	\cr
		\Gamma^{(1)i}_{jk}	&={2m\over r^3}\left( x_i-v_it \right)
                                  \delta_{jk} -{3m\over r^5}
                                  \left( x_ix_jx_k -x_ix_jv_kt
                                  -x_jx_kv_it-x_ix_kv_jt \right)\qquad .
                                                  	\cr}
         \eqno{\rm (41)}$$

{\bf (c) The Unperturbed Path}.
Let $t=0$ be the time of closest approach of the photon and lens,
and use co-ordinates centered on the lens with
co-ordinate axes chosen so that at $t=0$
the photon wavevector, $ \underline{n}_0$, points in the $y$
direction and the photon position vector, $\underline{r}_0$,
lies along the $z$ axis.
With this choice of co-ordinates the unperturbed path of the photon
is simply $x^{(0)}=(\lambda ,0,\lambda ,r_0 )$
where $r_0=\vert \underline{r}_0 \vert$ is a constant of the motion.

Analysis of the consistency criteria for our method quickly reveals
that we may work over the entire unperturbed path provided
$\epsilon^2\ll 1$ and $\epsilon^2\ll \kappa$. The first of these
inequalities limits us to regions of spacetime where the
Newtonian potential of the perturbing mass is small, and the second
restricts the mass to move at non-relativistic velocities. The second
of these conditions we have already imposed. The first amounts to
restricting our attention to weak lensing scenarios, that is, lensing
for which the impact parameter is much larger than twice the Schwarzschild
radius of the lens.

{\bf (d) The Bend Angle}.
We will call the deflection angle in the $yz$-plane the bend angle.
In the small angle approximation it is given by
$$\eqalign{	 \theta_{\rm bend}	&={dz \over dy}	\cr
				&= {{\dot z} \over {\dot y}}	\cr
				&= { {\dot z}^{(1)} \over
1+ {\dot y}^{(1)} }	\cr
				&\approx {\dot z}^{(1)}
= - \int_{ -\infty}^{+\infty} \left( \Gamma_{00}^{(1)3}
+2\Gamma_{02}^{(1)3}
+\Gamma_{22}^{(1)3} \right) d\lambda		\cr}\eqno{\rm (42)}$$
(using equation (15) in its first integral form, with $A
=B=0$).
Substituting the co-ordinate values on the unperturbed path into the
expressions for the connection coefficients, switching variables to
$s= \lambda /r_0$, and working to first order in
$v$ gives
$$ \theta_{\rm bend} =-{m\over r_0}\int_{-\infty}^{+\infty}
{ds \over \left( 1+s^2 \right) ^{5/2} } \left[ 3\left( 1-2v_y \right)
-v_zs +2v_zs^3 +{15s \left( v_z+sv_y \right) \over 1+s^2} \right]
	.\eqno{\rm (43)}$$
which may be integrated trivially to yield the result
$$ \theta_{\rm bend}=-{4m \over r_0} \left( 1-v_y \right)
\eqno{\rm (44)}$$
or, in three-vector form,
$$ \theta_{\rm bend}=-{4m \over r_0} \left( 1- \underline{v}\cdot
\underline{n}_0 \right)	.\eqno{\rm (45)}$$
which is valid to first order in $m/r_0$ and $v$.
This is the first result of our method and a short calculation
shows that
it agrees with a Lorentz transformation of the usual (static) lens
deflection angle to a frame in which the lens moves with velocity
$\underline{v}$.

{\bf (e) The Frequency Shift}.
The frequency shift of the photon between emission, $e$,
and observation, $o$, is defined by
$$\eqalign{	{\Delta\nu \over \nu}	&={\nu_o -\nu_e \over \nu_e}\cr
					&=(1+z)^{-1}-1		.\cr}
\eqno{\rm (46)}$$
where the redshift, $z$, is given by
$$(1+z)^{-1}={(k\cdot u )_o \over (k\cdot u)_e}.$$
with $u_o$ and $u_e$ the four-velocities of the observer and emitter,
respectively. Keeping
terms only to linear order,
$$(1+z)^{-1}= {k^{(0)}_{\mu}(o)u^{(0)\mu}(o) +k^{(0)}_{\mu}(o)
u^{(1)\mu}(o) +k^{(1)}_{\mu}(o)u^{(0)\mu}(o) \over
k^{(0)}_{\mu}(e)u^{(0)\mu}(e) + k^{(0)}_{\mu}(e)u^{(1)\mu}(e)
+ k^{(1)}_{\mu}(e)u^{(0)\mu}(e) } .\eqno{\rm (47)}$$
We take the observer and emitter to be far from the lens and at rest
relative to each other (letting the
observer and the emitter have some relative motion would merely give rise
to the usual Doppler terms). To be precise, the observer and emitter
four-velocities are parallel translates of each other along a path
passing far from the lens, this relationship being, at least
asymptotically, path-independent. This allows us to write
$u^{(0)}(o)=u^{(0)}(e)=(1,0,0,0)$. Furthermore, in the
asymptotic limit we are considering,
$u^{(1)}(o)=u^{(1)}(e)=(0,0,0,0)$. Making these
substitutions in equation (47) and combining the resulting expression with
equation (46) yields
$$ {\Delta \nu \over \nu}={k_0^{(0)}(o)+k_0^{(1)}(o) \over
k_0^{(0)}(e)+k_0^{(1)}(e) }-1 . \eqno{\rm (48)}$$
We showed in section II that for a perturbation of restricted scale
we could self-consistently assume that
the photon is unperturbed at emission, so that
$k^{(1)}(e)=0$. In addition, $k_0^{(0)}=g^{(0)}_{0\mu}k^{(0)\mu}
=-{\dot t}^{(0)}=-1$ at both emission and reception, from
which we deduce
$$\eqalign{ {\Delta \nu \over \nu}		&=-k_0^{(1)}(o) \cr
						&=k^{(1)0}(o)\equiv
	{\dot t}^{(1)}(o)	.\cr}\eqno{\rm (49)}$$

Taking our emission point to be $\lambda = -\infty$ and our reception
point to be $\lambda = +\infty$ equations (49) and (15) lead to
$${\Delta\nu \over \nu}=-\int^{+\infty}_{-\infty} \left(
\Gamma^{(1)0}_{00}+ 2\Gamma^{(1)0}_{02}+\Gamma^{(1)0}_{22} \right) d\lambda .
\eqno{\rm (50)}$$
Substituting in the co-ordinate values for the unperturbed path,
transforming variables to $s=\lambda /r_0 $, and performing a few
algebraic manipulations leads to
$${\Delta\nu \over \nu}=-{m \over r_0}\int_{-\infty}^{+\infty}
{ds \over \left[ 1-2sv_z+ \left( 1-2v_y \right)s^2 \right] ^{5/2} }
\left[ v_z+s \left( 2-3v_y \right) -6s^2v_z +2s^3 \left( 1-5v_y \right)
\right] ,\eqno{\rm (51)}$$
from which an expansion to ${\cal O}(v)$ yields
$${\Delta\nu \over \nu}=-{4m \over r_0}v_z \eqno{\rm (52)}$$
or, in three-vector notation
$${\Delta \nu \over \nu}=-{4m \over r_0^2} \underline{v}\cdot
\underline{r}_0 \qquad .\eqno{\rm (53)}$$
Like the formula for the bend angle, this result agrees with
previous calculations of the frequency shift (Birkinshaw and Gull,
1983): to ${\cal O}(v)$
the frequency shift arises from transverse motion of the lens.

{\bf (f) The Skew Angle}.
We will call the angular deflection out of the $yz$-plane the
skew angle. The special relativistic
treatment of the scattering of a photon
by a moving point mass
represents an instantaneous interaction between the photon and the
lens and so, necessarily, gives a vanishing skew angle.
Our method accounts for the interaction of the photon and the lens
over the entire photon path and confirms that the skew angle vanishes
by direct calculation to order $mv/r_0$.

In the small angle approximation the skew angle is given by
$$\eqalign{	\theta_{\rm skew}	&={dx \over dy}	\cr
				&={ {\dot x}\over {\dot y}}	\cr
				&={ {\dot x}^{(1)} \over
1+{\dot y}^{(1)} }	\cr
				&\approx {\dot x}^{(1)}
=-\int_{-\infty}^{+\infty} \left( \Gamma_{00}^{(1)1}
+2\Gamma_{02}^{(1)1}
+\Gamma_{22}^{(1)1} \right) d\lambda	.\cr} \eqno{\rm (54)}$$
which becomes
$$\eqalign{ \theta_{\rm skew}	&={mv_x \over r_0}\int_{-\infty}^{+\infty}
{s\, ds \over \left[ 1-2v_zs +\left( 1-2v_y \right) s^2 \right] ^{3/2} }\cr
	&\qquad -{3mv_x \over r_0}\int_{-\infty}^{+\infty} {s^3 \, ds \over
\left[ 1-2v_zs +\left(1-2v_y \right) s^2 \right] ^{5/2} } .\cr}
\eqno{\rm (55)}$$
which vanishes to first order in $v$.

\vskip0.25truein
\noindent
{\bf IV. Summary}

We have described a method for constructing null geodesics in
arbitrary metric perturbed spacetimes using null geodesics of the
background spacetime. The method fully generalizes the usual
Sachs-Wolfe technique for calculating temperature fluctuations of the
CMBR in metric perturbed spacetimes. Because our method constructs
both the spatial and timelike components of the perturbed geodesic
it is able to address questions of interest in gravitational lens
theory, such as the bend angles of the true path relative to the
unperturbed path. We have provided an explicit illustration with the
calculation of the behavior of a photon passing a moving lens.
A forthcoming paper will show how our method can be used to calculate
other quantities of gravitational lens theory, such as the
amplification undergone by a bundle of light rays passing a given
perturbation.

\vskip0.25truein
\noindent
{\bf Appendix A: Rewriting the Jacobi Operator}

We prove the equivalence of equations (15) and (18). This short
calculation has an interesting history. It is implicit in
Weinberg (1972), section 6.10. We produced the following proof in the course
of trying to gain a physical understanding of equation (15). After we
had done so, a comment in Burke (1985) drew our
attention to a paper by Faulkner and
Flannery (1978) where the same calculation appears in a different
context. It appears that this calculation is either obvious or not
depending on who is presenting it. It was not obvious to us, and we
feel its importance to the physical understanding of equation (15)
warrants its inclusion here.

Let $D/d\lambda$ denote covariant
differentiation along the curve $x^{(0)}$ with the background
connection. Then for an arbitrary vector $v$

$${D\over d\lambda}v^{\mu}={dv^{\mu}\over d\lambda}+\Gamma^{(0){\mu}}
_{\alpha\beta}
k^{(0)\alpha}v^{\beta}				\eqno{(A1)}	$$

\noindent
and

$$\eqalign{ {D^2\over d\lambda^2}v^{\mu}	&={d^2v^{\mu}\over d\lambda^2}
+\Gamma^{(0)\mu}_{\alpha\beta,\gamma}k^{(0)\alpha}
k^{(0)\gamma}v^{\beta}
	\cr
					&+\Gamma^{(0)\mu}_{\alpha\beta}
{dk^{(0)\alpha}\over d\lambda}v^{\beta}
+2\Gamma^{(0)\mu}_{\alpha\beta}k^{(0)\alpha}
{dv^{\beta}\over d\lambda}			\cr
					&+\Gamma^{(0)\mu}
_{\alpha\beta}k^{(0)\alpha}
\Gamma^{(0)\beta}_{\sigma\rho}k^{(0)\sigma}v^{\rho}.
		\cr}	\eqno{(A2)}$$

\noindent
Using the geodesic equation for $k^{(0)}$ this becomes

$$\eqalign{ {D^2\over d\lambda^2}v^{\mu}
	&={d^2v^{\mu}\over d\lambda^2}
+2\Gamma^{(0)\mu}_{\alpha\beta}k^{(0)\alpha}{dv^{\beta}\over d\lambda}	\cr
					&+\Gamma^{(0)\mu}
_{\alpha\beta ,\sigma}
k^{(0)\alpha}k^{(0)\sigma}v^{\beta}+\Gamma^{(0)\mu}
_{\alpha\beta}\Gamma^{(0)\alpha}_{\sigma\rho}k^{(0)\beta}
k^{(0)\sigma}v^{\rho}				\cr
					&-\Gamma^{(0)\mu}_{\alpha\beta}
\Gamma^{(0)\alpha}_{\sigma\rho}k^{(0)\sigma}k^{(0)\rho}v^{\beta}
	\cr}	\eqno{(A3)}$$

\noindent
The Riemann tensor of the background is given by

$$R^{(0)\mu}{}_{\alpha\beta\sigma}
=\left( \Gamma^{(0)\mu}_{\alpha\sigma ,\beta}
-\Gamma^{(0)\mu}_{\alpha\beta ,\sigma}
+\Gamma^{(0)\mu}_{\beta\rho}\Gamma^{(0)\rho}_{\alpha\sigma}
-\Gamma^{(0)\mu}_{\sigma\rho}\Gamma^{(0)\rho}
_{\alpha\beta} \right)	\eqno{(A4)}$$

\noindent
so that (A3) can be written

$$ {D^2\over d\lambda^2}v^{\mu}-R^{(0)\mu}
{}_{\alpha\beta\sigma}k^{(0)\alpha}k^{(0)\beta}v^{\sigma}
={d^2v^{\mu}\over d\lambda^2}+2\Gamma^{(0)\mu}
_{\alpha\beta}k^{(0)\alpha}{dv^{\beta}\over
d\lambda}+\Gamma^{(0)\mu}_{\alpha\beta ,\sigma}
k^{(0)\alpha}k^{(0)\beta}v^{\sigma} \eqno{(A5)}$$

\noindent
Comparing this equation to equation (15) completes the proof.

\vskip0.25truein
\noindent
{\bf Appendix B: The Path-Ordering }

This appendix provides a brief introduction to the use of the
path-ordering symbol. Recall that in section II we needed to solve a
matrix system of the form

$${\dot P}=MP		\eqno{(B1)}$$

\noindent
subject to $P(a,a)=1_d$.
If the quantities $P$ and $M$ were functions, the solution of this
system would be the usual exponential. An exponential with matrix
argument will {\it  not} work, however, for the reason that a matrix
function and its derivative matrix do not, in general, commute.

Rewriting equation (B1) as the
integral system

$$ P(\lambda ,a)=1_d+\int^{\lambda}_a M(\tau )P(\tau ,a)\, d\tau
				\eqno{(B2)}$$

\noindent
allows a possible solution, order by order in $M$,
to be written down by iteration,

$$P(\lambda ,a)=1_d+\int^{\lambda}_a M(\tau )\, d\tau
+\int^{\lambda}_a M(\tau )\, d\tau\, \int^{\tau}_a M(\tau^{\prime})
\, d\tau^{\prime}\, +\ldots		\eqno{(B3)}$$

\noindent
A simple proof of convergence then shows that this expression is, in
fact, a valid solution.
The region of integration for the n-th order term (in $M$) is known
as the n-simplex (the n-dimensional analog of the
triangle). We can extend the region of integration to the n-cube
in two steps. At each order we must symmetrize the n-fold
product $M(\tau )M(\tau^{\prime})\dots M(\tau^{\prime\ldots})$
completely in its arguments. This extends the
argument of the n-simplex integral to an argument defined over the
entire n-cube and ensures that each of the n!
simplices in the n-cube contributes the same amount to the total
integral over the n-cube, an amount equal to the
value of the original integral over the single simplex. Our second
step, then, is to divide each term by its overcounting factor, n! at
n-th order. These combinatorial factors yield the
exponential. Thus the path-ordering symbol amounts to a notice to
perform an integration written over a cube only over its ``lowest''
(in the sense of the discussion above) simplex.

\vskip0.25truein
\noindent
{\bf Appendix C: Another Transcription of the Solution }

In this appendix we present another form for the solution of the
perturbed Jacobi equation, needing only a single path-ordering but
containing terms whose geometrical meanings are more obscure than
those in equation (32).
We first write equation (15) as an eight-dimensional
first order system,

$$\pmatrix{ {\dot x}^{(1)} \cr {\dot k}^{(1)}\cr}=
\pmatrix{ 0&1_d\cr -B &-A \cr}\pmatrix{ x^{(1)}\cr k^{(1)} \cr}
+\pmatrix{ 0\cr f\cr}				\eqno{(C1)}$$

\noindent
where $1_d$ is a $4\times 4$ identity matrix and
$A$, $B$ and $f$ are given by equation (16). From this point
we will denote the $8\times 8$ matrix which is the coefficient matrix
of the associated homogeneous system by $M$.

We can now proceed to solve equation (C1) in exactly the same manner
as we solved equation (30) in section II.
We first obtain the associated transition matrix, $U$, which solves

$${d\over d\lambda}U(\lambda ,a)=M(\lambda)U(\lambda ,a) \eqno{(C2)}$$

\noindent
subject to $ U(a,a)=1_d$. We note that
the equation for $U$ is solved by a path-ordered exponential

$$U(\lambda ,a)={\cal P}\exp \left( \int_a^{\lambda}M(\tau)
\, d\tau \right)					\eqno{(C3)}$$

Having obtained $U$ it is easy to check by straightforward
differentiation that equation (C1) is solved by, with
$y:=(x^{(1)},k^{(1)})$ and $s:=(0,f)$,

$$y(\lambda)=U(\lambda ,a)y(a)+\int_a^{\lambda}U(\lambda ,\tau )
s(\tau )\, d\tau				\eqno{(C4)}$$

\noindent
This is the solution we desired to obtain.

The equivalence of this solution and that given in the main body of
the paper is easily established by working in the co-ordinate system
for which the Christoffel coefficients vanish along the unperturbed
path. The different appearance of the two solutions is essentially
equivalent to the two different ways of writing the Jacobi equation.
The usual expression for the Jacobi operator, involving covariant
derivatives and the Riemann tensor, has the advantage of being an
obviously geometrical quantity. Writing the Jacobi operator as the LHS
of equation (15) obscures its geometrical meaning but requires fewer
computations of coefficients.
The solutions mirror the strengths and
disadvantages of the two starting formulations although they are
completely equivalent.

\vskip0.25truein
\noindent
{\bf Appendix D: The Greens Function of Minkowski Space}

In this appendix
we construct the Greens function appropriate
to a perturbed Minkowski space and
show its equivalence to direct integration of equation (15).
In co-ordinates
for which the connection terms of the Minkowski background
vanish, the connector is the identity and
the transition matrix can be calculated instantly from equation (33)
with $R=0$, yielding

$$ U(\lambda , a)=
\pmatrix{ 1_d & (\lambda -a)1_d \cr 0 &1_d \cr}	\eqno{(D1)}$$

Simple matrix multiplication verifies that this solution satisfies
conditions (27) and (28) (rewritten with $U$ in place of $P$).
It is easy to see that the bottom four rows
yield a solution for the photon wavevector equivalent to simple
integration of the RHS of equation (15).
To see that the photon paths calculated by the
two techniques agree, start from
equation (32). Assuming vanishing initial data for convenience and using
the form of the propagator given in equation (D1) yields

$$ x^{(1)}(\lambda )=\lambda \int_a^{\lambda}s(\tau )\, d\tau
-\int_a^{\lambda }\tau s(\tau )\, d\tau		\eqno{(D2)}$$

\noindent
On the other hand, a direct second integration of equation (15) gives

$$x^{(1)}(\lambda )=\int_a^{\lambda }k^{(1)}(\tau )\, d\tau
=\int_a^{\lambda }\, d\tau \int_a^{\tau} s(\tau ')\, d\tau '
\eqno{(D3)}$$

\noindent
which can be seen to be equivalent to (D2) after integration by parts.

\vskip1.0truein
We would like to thank Sean Carroll for many helpful discussions, and
the referee, Eric Linder, for valuable comments on presentation and a
discussion of the consistency criteria problem.
This work was supported by the National Science Foundation under grant
AST90-05038.

\vfill\eject

\centerline{\bf References}

\vskip 12pt
\normalbaselineskip=8pt plus0pt minus0pt
                            \parskip 0pt
\def\ref#1  {\noindent \hangindent=24.0pt \hangafter=1 {#1} \par}

\ref{Anile, A.~M., \& Motta, S. 1976, ApJ, 207, 685}
%\ref{Anile, A.~M., and Motta, S. 1978, \mnras\vol{184} 319.}
\ref{Arg\"ueso, F., \& Martinez-Gonzalez, E. 1989, MNRAS, 238, 1431}
\ref{Arg\"ueso, F., Martinez-Gonzalez, E., \& Sanz, J.~L. 1989, ApJ,
336, 69}
%\ref{Bardeen, J., Steinhardt, P., and Turner, M. 1983,
%{\sl Phys. Rev.} {\bf D28}, 679.}
%\ref{Barrow, J.~D., Juszkiewicz, R., and Sonoda, D.~H. 1983, \nat
%\vol{305} 397.}
\ref{Birkinshaw, M., \& Gull, S.~F. 1983, Nature, 302, 315}
\ref{Bond, J.~R., \& Efstathiou, G. 1984, ApJ, 285, L45}
\ref{Bond, J.~R., \& Efstathiou, G. 1987, MNRAS, 226, 655}
\ref{Burke, W.~L. 1985, Applied Differential Geometry,
Cambridge University Press, Cambridge}
%\ref{Collins, C.~B., and Hawking, S.~W. 1973, \mnras\vol{162} 307.}
\ref{De Felice, F., \& Clarke, C.~J.~S. 1990, Relativity on
Curved Manifolds, Cambridge University Press, Cambridge}
%\ref{Dyer, C.~C. 1976, \mnras\vol{175} 429.}
\ref{Ehlers, J., Geren, P., \& Sachs, R.~K. 1968, J. Math. Phys.,
9, 1344}
%\ref{Fabbri, R., Guidi, I., and Natale, V. 1982, \apj\vol{257} 17.}
%\ref{Faraoni, V. 1992, \apj {\sl to be published} }
\ref{Faulkner, J., \& Flannery, B.~P. 1978, ApJ, 220, 1125}
\ref{Futamase, T. 1988, Phys. Rev. Lett., 61, 2175}
\ref{Futamase, T. 1989, MNRAS, 237, 187}
\ref{Futamase, T., \& Sasaki, M. 1989, Phys. Rev. D, 40, 2502}
%\ref{Goicoechea, L.~J., and Sanz, J.~L. 1984, {\sl Phys. Rev. D{\bf 29}},
%607.}
%\ref{Goicoechea, L.~J., and Sanz, J.~L. 1985, \apj\vol{293} 17.}
\ref{Grishchuk, L.~P., \& Zeldovich, Ya.~B. 1978, Soviet
Astronomy-AJ, 22, 125}
%\ref{Guth, A.~H., and Pi, S.~Y. 1982, {\sl Phys. Rev. Lett.} {\bf 49},
%1110.}
%\ref{Guth, A.~H., and Pi, S.~Y. 1985, {\sl Phys. Rev. D{\bf 32}},
%1899.}
%\ref{Hawking, S. 1982, {\sl Phys. Lett.} {\bf 115B}, 295.}
\ref{Humi, M., \& Miller, W. 1988, Second Course in Ordinary
Differential Equations for Scientists and Engineers, Springer-Verlag,
New York}
%\ref{Hawking, S. 1969, \mnras\vol{142} 129.}
%\ref{Kaiser, N. 1982, \mnras\vol{198} 1033.}
%\ref{Kristian, J., and Sachs, R.~K. 1966, \apj\vol{143} 379.}
\ref{Jacobs, M.~W., Linder, E.~V., \& Wagoner, R.~V. 1992,
Phys. Rev. D, 45, R3292}
\ref{Jordan, P., Ehlers, J., \& Sachs, R.~K. 1961, Akad.
Wiss. Lit. Mainz Abh. Math.-Nat. Kl., 1, 3}
%\ref{Lifshitz, E.~M. 1946, {\sl J. Phys. U.S.S.R.},\vol{10} 116.}
\ref{Linder, E.~V. 1988a, ApJ, 326, 517}
\ref{Linder, E~.V. 1988b, ApJ, 328, 77}
%\ref{Linder, E.~V. 1988c, \aa\vol{206} 199.}
\ref{Linder, E~.V. 1990, MNRAS, 243, 362}
\ref{Martinez-Gonzalez, E., Sanz, J.~L., \& Silk, J. 1990,
ApJ, 355, L5}
%\ref{Nottale, L. 1984, \mnras\vol{206} 713.}
%\ref{Novikov, I.~D. 1968, {\sl Soviet Astronomy-AJ},\vol{12} 427.}
\ref{Peebles, P.~J.~E., \& Yu, J.~T. 1970, ApJ, 162, 815}
%\ref{Raine, D.~J., and Thomas, E.~G. 1981, \mnras\vol{195} 649.}
%\ref{Rees, M.~J., and Sciama, D.~W. 1968, \nat\vol{217} 511.}
\ref{Sachs, R.~K. 1961, Proc. Roy. Soc. A, 264, 309}
\ref{Sachs, R.~K., \& Wolfe, A.~M. 1967, ApJ, 147, 73}
%\ref{Sasaki, M. 1987, \mnras\vol{228} 653.}
%\ref{Sanz, J.~L. 1979, {\sl Phys. Rev. D{\bf 20}}, 1791.}
%\ref{Smoot, G.~F., {\sl et. al.} 1992, \apjl
%{\sl to be published.}}
%\ref{Starobinsky, A. 1982, {\sl Phys.Lett.} {\bf 117B}, 175.}
\ref{Suto, Y., Gouda, N., \& Sugiyama, N. 1990, ApJS, 74, 665}
\ref{Synge, J.~L. 1960, Relativity: The General Theory,
North-Holland, Amsterdam}
%\ref{Vittorio, N., and Silk, J. 1984, \apjl\vol{285} L39.}
%\ref{Vittorio, N., and Silk, J. 1992, \apjl\vol{385} L9.}
\ref{Watanabe, K., \& Tomita, K. 1990, ApJ, 355, 1}
\ref{Weinberg, S. 1972, Gravitation and Cosmology, John Wiley
and Sons, New York}
%\ref{Wright, E.~L., {\sl et. al.} 1992, \apjl {\sl to be published.}}
%
\normalbaselineskip=24pt plus0pt minus0pt
                  \parskip 12.0pt
\vfill\eject

\end